Perspective

# Strategic priorities for transformative progress in advancing biology with proteomics and artificial intelligence

Yingying Sun[1,2], Jun A[1,2], Zhiwei Liu[1,2], Rui Sun[1,2], Liujia Qian[1,2], Samuel H. Payne[3], Wout Bittremieux[4], Markus Ralser[5], Chen Li[6], Yi Chen[1,2], Zhen Dong[1,2], Yasset Perez-Riverol[7], Asif Khan[8], Chris Sander[9], Ruedi Aebersold[10], Juan Antonio Vizcaíno[7], Jonathan R Krieger[11], Jianhua Yao[12], Han Wen[13], Linfeng Zhang[13], Yunping Zhu[14], Yue Xuan[15], Benjamin Boyang Sun[16], Liang Qiao[17], Henning Hermjakob[7], Haixu Tang[18], Huanhuan Gao[2], Yamin Deng[2], Qing Zhong[19], Cheng Chang[14], Nuno Bandeira[20], Ming Li[21], Weinan E[22], Siqi Sun[23], Yuedong Yang[24], Gilbert S. Omenn[25], Yue Zhang[26], Ping Xu[14], Yan Fu[27], Xiaowen Liu[28], Christopher M. Overall[29], Yu Wang[30], Eric W. Deutsch[31], Luonan Chen[32], Jürgen Cox[33], Vadim Demichev[5], Fuchu He[14,34], Jiaxing Huang[26], Huilin Jin[35], Chao Liu[36], Nan Li[37], Zhongzhi Luan[38], Jiangning Song[6], Kaicheng Yu[26], Wanggen Wan[39], Tai Wang[40], Kang Zhang[41], Le Zhang[42], Peter A. Bell[43], Matthias Mann[44]*, Bing Zhang[45]*, Tiannan Guo[1,2]*

1, Affiliated Hangzhou First People's Hospital, State Key Laboratory of Medical Proteomics, School of Medicine, Westlake University, Hangzhou, Zhejiang Province, China
2, Westlake Center for Intelligent Proteomics, Westlake Laboratory of Life Sciences and Biomedicine, Hangzhou, Zhejiang Province, China
3, Biology Department, Brigham Young University, Provo, Utah 84602, United States
4, Department of Computer Science, University of Antwerp, 2020 Antwerp, Belgium
5, Department of Biochemistry, Charité Universitätsmedizin Berlin, Berlin, Germany.
6, Biomedicine Discovery Institute and Department of Biochemistry and Molecular Biology, Monash University, Melbourne, Victoria, VIC 3800, Australia
7, European Molecular Biology Laboratory, European Bioinformatics Institute (EMBL-EBI), Wellcome Genome Campus, Hinxton, Cambridge, CB10 1SD, UK
8, Harvard Medical School, Ludwig Center at Harvard
9, Harvard Medical School, Broad Institute, Ludwig Center at Harvard, Dana-Farber Cancer Institute
10, Department of Biology, Institute of Molecular Systems Biology, ETH Zürich, Zürich, Switzerland
11, Bruker Ltd., Milton L9T 6P4, Ontario, Canada
12, AI for Life Sciences Lab, Tencent, Shenzhen, 518057, China
13, State Key Laboratory of Medical Proteomics, AI for Science Institute, 100080, Beijing, China
14, State Key Laboratory of Medical Proteomics, Beijing Proteome Research Center, National Center for Protein Sciences(Beijing), Beijing Institute of Lifeomics, Beijing 102206, China
15, Thermo Fisher Scientific GmbH, Hanna-Kunath Str. 11, Bremen, 28199, Germany
16, Informatics and Predictive Sciences, Research, Bristol Myers Squibb, US
17, Department of Chemistry, Fudan University, Songhu Road 2005, Shanghai 200438, China
18, Department of Computer Science, Luddy School of Informatics, Computing and Engineering, Indiana University, IN 47408, United States
19, ProCan®, Children's Medical Research Institute, Faculty of Medicine and Health, The

University of Sydney, Westmead, NSW, Australia
20, Center for Computational Mass Spectrometry, Dept. Computer Science and Engineering, Skaggs School of Pharmacy and Pharmaceutical Sciences, University of California, San Diego (UCSD), La Jolla, California, USA
21, Central China Institute of Artificial Intelligence and University of Waterloo
22, AI for Science Institute, Center for Machine Learning Research, School of Mathematical Sciences, Peking University, P. R. China
23, Research Institute of Intelligent Complex Systems, Fudan University, Shanghai, China
24, School of Computer Science and Engineering, Sun Yat-sen University, Guangzhou, China
25, Departments of Computational Medicine & Bioinformatics, Internal Medicine, Human Genetics, and Environmental Health
26, School of Engineering, Westlake University, Hangzhou, Zhejiang 310024, China
27, Academy of Mathematics and Systems Science, Chinese Academy of Sciences, Beijing 100190, China
28, Deming Department of Medicine, School of Medicine, Tulane University, New Orleans, LA, USA
29, Department of Oral Biological and Medical Sciences, Centre for Blood Research, University of British Columbia, Vancouver, British Columbia V6T 1Z4, Canada
30, Pengcheng Laboratory, Shenzhen, 518055, China
31, Institute for Systems Biology (ISB), Seattle, Washington, USA
32, Key Laboratory of Systems Health Science of Zhejiang Province, Hangzhou Institute for Advanced Study, University of Chinese Academy of Sciences, Chinese Academy of Sciences
33, Computational Systems Biochemistry Research Group, Max Planck Institute of Biochemistry, Am Klopferspitz 18, 82152 Martinsried, Germany
34, International Academy of Phronesis Medicine (Guang Dong), No. 96 Xindao Ring South Road, Guangzhou International Bio Island, Guangzhou, 510000, China
35, School of Big Data, Anhui University, Hefei, China
36, School of Biological Science and Medical Engineering & School of Engineering Medicine, Beihang University, Beijing, China
37, Westlake University High-Performance Computing Center, Westlake University, Hangzhou, Zhejiang, China
38, School of Computer Science and Engineering, Beihang University, Beijing, 100191, China
39, School of Communication and Information Engineering, Institute of Smart City, Shanghai University, Shanghai, China
40, 250 Water St, Bristol Myers Squibb, Cambridge, 02141 USA
41, Eye Hospital and Institute for Advanced Study on Eye Health and Diseases, Institute for clinical Data Science, Wenzhou Medical University, Wenzhou, China
42, Department of Computer Science, Sichuan University, Chengdu, China
43, Department of Oral Biological and Medical Sciences, University of British Columbia, Vancouver, BC V6T 1Z3, Canada
44, Department Proteomics and Signal Transduction, Max Planck Institute of Biochemistry, Martinsried, Germany
45, Lester and Sue Smith Breast Center, Baylor College of Medicine, Houston, TX 77030, USA

**Abstract**

Artificial intelligence (AI) is transforming scientific research, including proteomics. Advances in mass spectrometry (MS)-based proteomics data quality, diversity, and scale, combined with groundbreaking AI techniques, are unlocking new challenges and opportunities in biological discovery. Here, we highlight key areas where AI is driving innovation, from data analysis to new biological insights. These include developing an AI-friendly ecosystem for proteomics data generation, sharing, and analysis; improving peptide and protein identification and quantification; characterizing protein-protein interactions and protein complexes; advancing spatial and perturbation proteomics; integrating multi-omics data; and ultimately enabling AI-empowered virtual cells.

**Introduction**

Molecular biology is highly complex and dynamic, and our understanding is still rudimentary, making it challenging to predict even a single cell's behavior. Advancing research and developing predictive models are essential for biomedical progress, ultimately enhancing disease diagnosis, treatment, and prevention. Achieving these goals relies on a comprehensive characterization of cellular molecular states, particularly in response to environmental changes and perturbations.

As the primary functional molecules of life, proteins orchestrate cellular functions, including cellular structure, signaling, metabolism, and growth regulation[1]. They are key players in health and disease, serving as critical drug targets and therapeutic agents[2]. Understanding cellular states goes beyond protein identification, as function depends on post-translational modifications (PTMs), subcellular localization, and binding partners, among other factors. Mass spectrometry (MS)-based proteomics is essential for identifying and quantifying proteins while revealing PTMs, single amino acid variants (SAAVs), isoforms, and protein-protein interactions (PPIs). It also uncovers protein complexes and structural changes in complex cellular backgrounds[3]. With advances in sensitivity and throughput, emerging fields such as single-cell proteomics, spatial proteomics, and perturbation proteomics are revolutionizing our understanding of biological organization and dynamics.

Proteomics has long used AI to enhance peptide and protein identification, such as predicting chromatographic retention time, ion mobility, and peptide fragmentation. Semi-supervised learning in modern search engines now accurately distinguishes true from false spectral annotations[4,5]. These advances have expanded the scope of proteomics, enabling the identification of proteinogenic peptides, antibodies, neoantigens, immune peptides, neuropeptides, and natural products[6]. AI has emerged as a key driver of clinical proteomics, aiding the discovery of biomarkers and drug targets[7,8]. These breakthroughs are driving transformative progress, yielding valuable biological and clinical insights.

Although AI has transformed proteomics data analysis, its broader application to ambitious biological goals remains challenging. Vast datasets, containing millions of MS raw files as of January 2025, are stored in public repositories like PRIDE[9] and MassIVE[10], yet only a small fraction has been used for AI training to date. The primary obstacle is the lack of an AI-friendly ecosystem for proteomics data generation, sharing, and analysis. Challenges arise at the data generation stage due to diverse sample preprocessing, liquid chromatography (LC), and MS methods, resulting in data with varying overall experimental approaches, formats, quality, and batch effects. Furthermore, differences in mass spectrometers, MS techniques, and quantification strategies impede data standardization and normalization. This is particularly evident in human proteome datasets, where small, disease-focused cohorts must be integrated with larger population datasets for AI-driven disease and therapy predictions. Acquiring time-series data and high-quality paired metadata remains a significant challenge. The field urgently needs expanded benchmark datasets for robust model validation. Beyond data challenges, achieving AI's full potential in proteomics necessitates clear research goals, optimized algorithms, enhanced model transparency and interpretability, and stronger interdisciplinary collaboration.

To accelerate progress in this field, we must define clear AI objectives and develop high-quality, large-scale, AI-friendly proteomic datasets. ImageNet[11] (for images) and PDB[12] (for protein structures) exemplify how structured data can drive AI innovations, and how the inherent complexity of these data can be simplified and made accessible to AI practitioners. Here, we outline two fundamental requirements for AI in proteomics: sufficient AI-friendly training data and an optimized ecosystem. Additionally, we identify six key areas where AI can be leveraged to advance data analysis and biological discovery, calling for global collaboration between AI and proteomics researchers. International initiatives like the π-HuB project [13], the Human Proteome Project (HPP) [14], and the International Cancer Proteogenome Consortium (ICPC)[15] are poised to play a crucial role in accelerating progress.

**Key area 1: development of an AI-friendly ecosystem for proteomics data**

The foundation of successful AI implementation in proteomics rests on training datasets that are easily accessible, large enough for model generalization, and high-quality to ensure robust performance.

*Data generation*

Building an AI-friendly proteomics ecosystem faces fundamental obstacles starting from data generation. MS data quality varies due to variability in sample preparation (*e.g.*, various sample processing methods), LC (*e.g.*, various LC methods, inadvertent contamination), and MS instrument performance (*e.g.*, various MS methods, vendors, environmental fluctuations, contamination accumulates)[16,17]. These variabilities compound exponentially as the data sets' sizes and sources expand.

The integration of standard operating procedures (SOPs), quality control (QC), and AI offers a powerful approach to mitigate data variability and enhance data quality. Harmonized MS platforms and standardized acquisition enhance reproducibility across laboratories [18,19]. While universal adoption of standard procedures is unrealistic, they remain essential for AI training datasets. Dedicated facilities like the Clinical Proteomics Tumor Analysis Consortium (CPTAC)[20], ProCan[21], Biognosys, and possibly π-HuB[13] can operate multiple instruments in a controlled, standardized manner to generate AI-suitable data.

Quality control represents a critical cornerstone in managing experimental variability. However, a universal QC method is still lacking, and QC measurements are subject to the same variations in sample preparation, LC, and MS performance as experimental samples. Recent work has proposed universal QC using synthetic data tailored to samples and instruments [22], whose implementation would inevitably rely on AI. Machine learning demonstrated success in selecting peptide precursors and predicting LC-MS performance in DIA-based QC workflows [23]. To further advance the field, SOP data should be integrated into AI models to improve model objectivity, while AI-driven simulations of experimental methods can optimize SOPs.

A critical challenge is the lack of a unified method to assess MS raw data quality. Integrating metrics like peak intensity, single to noise ratio (SNR), retention time reproducibility, and chromatographic resolution would enable automated evaluation of large datasets, thereby ensuring consistent filtering of low-quality data for machine learning training.

*Data synthesis*

Synthetic data offers an essential complement for AI modeling, particularly for benchmarking, algorithm evaluation, and experimental design optimization[24]. It enables the generation of sample-specific data tailored to processing methods, separation techniques, and acquisition parameters, without relying on prior analyses. By leveraging machine learning to replace empirical approaches, synthetic data enables refinement of LC-MS/MS acquisition parameters, leading to optimized experimental designs[22]. In cases, where real training data is insufficient, synthetic data can serve as a crucial resource [25]. Recently, simulated datasets have enhanced DIA-MS software through pretrained models, ensuring accurate peak area calculations in training sets[26].

*Raw MS data format*

MS raw data types are highly diverse due to variations in instruments, MS acquisition methods (*e.g.*, DDA, DIA, SWATH, PASEF), and labeling techniques (*e.g.*, TMT, iTRAQ, SILAC). While standard open data formats like mzML, mzIdentML, mzTab, and mzQC developed by HUPO Proteomics Standards Initiative (HUPO-PSI)[27], enable spectral data sharing, but they are designed for single runs, not large-scale aggregation. Unlike text, images or audio, but similarly to other scientific areas, proteomics data requires domain expertise and specialized software to interpret, creating significant barriers for computer scientists. Advancing AI-driven methods, demand new data structures based on formats like TensorFlow [28] and Parquet [29] for large-scale data sharing and processing. This transformation necessitates collaboration between MS vendors and data engineers to adopt unified, user-readable data types for building AI-friendly datasets.

*Metadata*

Metadata plays a central role in diverse AI applications within proteomics. These include classification tasks like sample classification (*e.g.*, cancer or non-cancer, tissue types) and protein function prediction, regression tasks such as predicting peptide properties (*e.g.*, retention time), and *de novo* sequencing tasks that translate spectra into peptide sequences. Proteomics metadata encompasses sample labels (*e.g.*, clinical phenotype, sample types, species) and proteomics-specific labels (*e.g.*, quantification methods, MS acquisition methods, instrument types, and batch design). While the SDRF-Proteomics format was designed to store and link metadata with raw data [30], its adoption in proteomics has been limited so far [31]. Additionally, the limited use of laboratory information management systems (LIMS) has hindered proper data structuring before being made publicly available. Furthermore, as with other omics fields, sharing metadata, especially clinical data, presents substantial hurdles due to ethical concerns. Federated learning offers a potential solution, enabling local model training on simulated sites behind firewalls and aggregating them into a global model [32].

*Data storage and sharing*

The accessibility of AI-friendly raw data and metadata in through public repositories, such as ProteomeXchange[31] forms the backbone of effective data sharing. Creating AI training datasets tailored to specific tasks, as outlined in the following sections, represents another critical requirement. A critical issue with data exchange platforms like ProteomXchange is frequent lack of defining the content and purpose of each file, which limits the usability of public datasets. To address this, we should either encourage repositories to require a table explaining each file's content and use in the publication or establish a community-wide standard. Beyond MS-based proteomics, affinity-based data like antibody-based approaches[33] and Olink data[34] demand collaborative efforts across disciplines for AI-friendly generation, analysis, storage, and sharing.

*Raw data interpretation*

Identifying and quantifying peptides and proteins from raw data underpins most proteomics AI training tasks. While conventional proteomics software suffices for small-scale data interpretation, large-scale datasets demand more sophisticated approaches. To meet this need, community resources and pipelines such as MassIVE.quant[35] and quantms[36] now provide annotated and reanalyzed datasets, enabling researchers to access and utilize large, curated collections of data. Future developments should prioritize using these tools to expand AI training datasets.

Another major challenge is standardizing quantitative results across diverse quantification methods (*e.g.*, TMT, iTRAQ, SILAC, label-free) and affinity-based data. Despite the availability of several normalization tools, a standardized workflow has yet to emerge. The details of these issues will be elaborated in depth in the second key area.

*Benchmarking datasets*

Universal benchmarking datasets play a vital role in evaluating software tool performance. Proteobench[37] offers a comprehensive framework for benchmarking identification and quantification based on DDA-MS and DIA-MS using database search or *de novo* peptide sequencing. ProteomicsML[38] delivers pre-packaged datasets and tutorials for training and validation of ML models in proteomics. Moreover, large-scale synthetic peptide datasets, such as proteomeTools[39] and peptide phosphorylated counterpart libraries[40], have been used to assess the performance of tools like *de novo* sequencing, phosphorylation-site localization, and fragmentation methods. The expansion of benchmarking datasets across species, MS instruments, acquisition methods, and diverse tasks like protein complex identification and spatial proteomics would catalyze advancement across the field.

*Building an AI-friendly ecosystem for proteomics*

The development of standardized AI-friendly data formats and workflows enables both the creation of new large-scale proteomics datasets and the transformation of existing public repository data into AI-ready resources. Following the successful examples of WordNet[41] and ImageNet[11] in other fields, these curated resources would advance AI in proteomics and promote standardized data generation through mature SOPs, fostering an AI-friendly ecosystem for data generation, sharing, and analysis.

The organization of community competitions where AI approaches can be used, like the Critical Assessment of Structure Prediction (CASP) [42], would inspire collaboration between AI and proteomics experts to address key challenges. The exponential growth in dataset size and complexity demands significantly greater computational power for AI model development, highlighting the urgent need for support from computing infrastructure providers.

**Key area 2: identification and quantitation of peptides and proteins**

The core task in MS-based proteomics is peptide and protein identification and quantification. Here, we highlight AI-driven advances, current challenges, and future solutions for AI integration, with a focus on large cohort studies, single-cell proteomics, plasma proteomics, and metaproteomics.

*Database search in bottom-up proteomics*

In DDA-MS-based bottom-up proteomics, the spectrum-centric database search strategy matches each spectrum to a peptide. Subsequently, peptide-to-spectrum matches (PSMs) are scored and filtered for false positives using a target-decoy approach. Deep learning-based PSM rescoring tools have greatly enhanced peptide identification[43]. The transformer-based DDA-BERT model performs well across mass spectrometers, single-cell proteomes, and multi-species datasets [44]. Recently, DeepSearch pioneered a modified transformer to rank PSMs based on cosine similarity between spectrum and peptide embeddings, offering a novel AI-driven approach for database searching. [45]. Additionally, emerging deep learning-based full-spectrum prediction methods promise to enhance identification rates[46].

DIA-MS generates highly convoluted spectra, which can be searched using both spectrum-centric and peptide-centric methods, with tools like DIA-NN[47], MaxDIA[48], and Spectronaut[49]. These tools are at the forefront of the AI revolution in proteomics. DIA-NN[47], which uses deep neural networks to assign $q$-values, has become a standard AI-based tool for DIA-MS analysis, though it is now closed-source. Newer open-source tools like AlphaPeptDeep[50], built based on the PyTorch DL library, match or surpass existing tools in predicting peptide properties. Recently, AlphaDIA[51], an end-to-end transfer learning tool for feature-free proteomics data analysis, has been developed. The newly developed tool, DIA-BERT[26], also leverages pretrained large language models to analyze the DIA data.

However, several critical challenges remain. i) Deep learning-based re-scoring often requires models trained individually for each MS file, increasing the risk of overfitting[26]. ii) Few AI models have been designed for the identification and quantification of proteoforms (*e.g.*, SAAVs, isoforms, PTMs). iii) Models require enhanced compatibility across species, MS instruments (e.g., Orbitrap, timsTOF, Astral), fragmentation methods (*e.g.*, HCD, CID, ETD, ECD), and peptide quantification labels (*e.g.*, TMT, SILAC, iTRAQ). iv) Many spectra remain uninterpreted[52,53], and spectra are often chimeric, requiring algorithms to identify multiple peptide precursors from a single spectrum[54]. v) Significant efforts are still needed to improve the identification rate, accuracy, generalization, and interpretability of these AI-based tools. vi) The integration of AI-driven intelligent data acquisition enables real-time database searching [55].

*De novo sequencing in bottom-up proteomics*

*De novo* peptide sequencing has advanced the identification of proteoforms (*e.g.*, SAAVs, isoforms, PTMs), small proteins (encoded by sORFs), "understudied proteins (USPs)", and other novel sequences beyond current knowledge. Deep learning tools have enhanced the identification of peptides such as immune peptides, antibodies, neoantigens, and bioactive peptides[6]. Furthermore, deep learning combined with multiple mirror proteases is shown to significantly improve sequence coverage and confidence[56].

However, *de novo* sequencing generally has lower identification rates than database search strategies while accuracy, precision, and recall of peptides still require improvement. As always in deep learning, overfitting or memorizing the dataset of interest (e.g. the human proteome), must be avoided. While algorithms can generalize well to known distributions, their performance drops significantly when generalizing to unknown, out-of-distribution (OOD) data. This is particularly important for *de* *novo* sequencing tasks, which rely on algorithms to uncover new knowledge. A promising solution is enabling models to reason. For example, recent advancements in large language models (LLMs) like GPT-o1 and DeepSeek-R1[57] allow chain-of-thought reasoning to tackle unseen challenges, demonstrating better generalization than traditional models like GPT-4[58] and DeepSeek-v3[59]. To train such models, data on the reasoning process—where humans provide both labels and rationale—can be valuable. Literature also serves as an additional data source, though current LLM and retrieval-augmented generation (RAG) technologies already address this. Additionally, current methods prioritize overall peptide or protein identification, while the explanation or annotation of why ions and signals match are often overlooked in the database search algorithm. Post-annotation of PSMs promise to improve the interpretability of large-scale data processing and uncover novel findings in MS experiments or biomedical applications.

*Top-down proteomics*

Top-down proteomics (TDP) analyzes intact proteins without digestion, enabling the detection of proteoforms (*e.g.*, isoforms, PTMs) even at low abundance and providing direct quantification[60]. However, TDP faces hurdles with MS data interpretation due to broader charge distributions and more product ions, while a smaller research community, fewer software tools, and limited datasets constrain AI applications in this field. As TDP attracts continued attention and AI advances, AI-driven approaches may transform the field.

AI-driven spectrum clustering tools like GLEAMS[52] and DLEAMSE[61], demonstrate potential for detecting the "dark proteome" by annotating unidentified mass spectra. Emerging methods such as COrrelation-based functional ProteoForm (COFP)[62] and inference of peptidoforms (IPF)[63] are advancing proteoform detection, and AI integration could further improve proteoform and USP characterization.

*AI enhanced emerging applications*

AI innovations are transforming emerging proteomics fields—such as studies involving large cohorts, single-cell, plasma, and metaproteomics.

*Large cohort studies.* Large cohort studies grappling with missing values and batch effects. Traditional proteomics software employ methods like match between run (MBR)[64] for imputation and Combat[65] and HarmonizR[66] for batch correction, though their accuracy and effectiveness remain areas for improvement[67]. Recently, deep learning tools like BERNN[68], DeepRTAlign[69], demonstrate advances in batch effect correction. Continued AI developments will further address these challenges.

*Single-cell proteomics (SCP).* SCP offers unprecedented insights into biological heterogeneity but struggles with low MS signal intensity, poor signal-to-noise ratio (SNR), sparse quantification, and low throughput[70]. AI-driven approaches to enhance low-intensity MS signals and reduce noise are crucial for improving SCP identification and quantification accuracy. Additionally, SCP faces analogous challenges to those found in large-cohort studies, due to the large number of samples.

*Plasma proteomics.* Plasma proteomics enables disease screening and biomarker discovery[71-73] but faces challenges like high dynamic range, reproducibility issues, batch effects, and data variability. AI can greatly enhance data acquisition and analysis, propelling plasma proteomics beyond traditional approaches like low-abundance protein enrichment, or high-abundance protein depletion.

*Metaproteomics.* Metaproteomics confronts high PSM mismatch rates, protein inference challenges, and especially large and not fully characterized search spaces[74]. AI-driven solutions like *de novo* sequencing[75,76], deep learning-based PSM re-scoring, and homolog recognition[77] enhance accuracy, though adoption remains limited. Efficient, lightweight models are essential to balance performance and computational cost.

Critically, more focus is needed on AI model transparency and interpretability, particularly for clinical applications such as disease prediction and drug target discovery.

**Key area 3: identification and quantification of protein complexes, protein-protein interactions, and other molecular interactions**

Most proteins function within complexes, and those involved in the same cellular processes frequently interact[78]. PPIs and protein complexes are highly dynamic and context-specific[3,79-81]. Identifying and quantifying these interactions and their perturbations across conditions fundamentally advances our understanding of biological processes and their roles in disease. Beyond protein-protein interactions, these molecular networks encompass interactions with small molecules like drugs, nucleic acids, and lipids.

The convergence of MS-based experiments and AI have transformed PPI and protein complex identification and quantification. Affinity purification MS serves as a cornerstone method[82] for PPI identification, and has recently been pushed to very high throughput with unprecedented data quality[83]. Its integration with deep learning has enhanced resolution of prefoldin holo- and subcomplex variants, complex–complex interactions, and complex isoforms[84]. Co-fractionation MS (CF-MS), including size exclusion chromatography-MS and blue native electrophoresis, is widely used for protein complex identification[85,86]. AI-based approaches can further refine protein complex identification and quantify abundance and composition changes across conditions[84,87-89]. Cross-linking MS (XL-MS) can be used to identify protein-protein and protein-RNA interfaces by combining cross-linking reactions, revealing residue proximity and complex topology [90]. AlphaLink [91] integrates XL-MS data with AlphaFold2's pair representation, significantly improving PPI predictions[92]. Its latest version, AlphaLink2[93], incorporates experimental constraints into AlphaFold-Multimer, extending predictions to protein complexes[94]. Beyond MS-based methods, structural biology techniques complement PPI and protein complex studies. AlphaFold 3 leverages deep learning to predict the joint structures of complexes, including proteins, nucleic acids, small molecules, ions, and modified residues[95].

Despite the growing accumulation of data from the above technologies, integrating diverse datasets to build large-scale, high-quality, AI-friendly protein complex databases remains a major challenge.

i) Data heterogeneity and format inconsistencies. Researchers generate protein complex data through various sources, including MS-based methods (*e.g.*, AP-MS, CF-MS, XL-MS), molecular biology experiments (*e.g.*, co-immunoprecipitation), manual curation and ML-based prediction (*e.g.*, Complex Portal[96]), bioinformatics predictions (*e.g.*, PPI databases[97]), and structural techniques (*e.g.*, X-ray crystallography, cryo-electron microscopy, and nuclear magnetic resonance). Differences in format, resolution, and coverage render normalization and integration difficult, limiting their direct applicability to AI models.

ii) Lack of standardized definitions and protocols. Inconsistent definitions of protein complexes across PPI databases and the absence of standardized detection protocols hinder data integration, reproducibility, and comparability.

iii) Dynamic and heterogeneous nature of protein complexes. Protein complexes exhibit complex dynamics, with temporal and spatial variations that static datasets often fail to capture. Low-abundance interactions add noise, making accurate detection more challenging.

iv) AI-driven data integration and QC. AI-based tools can play a crucial role in normalizing, integrating, and assessing data quality to ensure consistency and reliability. Standardized protocols and optimized AI models are fundamental to advancing protein complex research.

**Key area 4: spatial proteomics**

Spatial proteomics, which maps protein localization and dynamics within cells and tissues, fundamentally advances our understanding of protein functions, interactions, cellular processes, and disease mechanisms. It consists of two layers: one focusing on different tissue regions and the other on cellular compartments or subcellular localizations.

Recently, deep visual proteomics (DVP)[98] emerged to enable single-cell analysis of tissue samples. It revolutionizes spatial proteomics by identifying cell-type-specific markers in stained cells and has already led to life-saving treatments[99]. Another method, filter-aided expansion proteomics (FAXP)[100], demonstrates strong reproducibility and identification depth in clinical formalin-fixed, paraffin-embedded (FFPE) samples. Furthermore, spatial proteomics technology based on immunofluorescence (IF) microscopy has generated over 80,000 confocal images, mapping substantial proteins across 30 cellular compartments and substructures[11].

Deep learning has revolutionized the scalable annotation of single-cell protein distributions in these images, opening new horizons for machine learning to identify cellular phenotypes, isolating nuclei, mapping protein subcellular localizations, and advancing spatial proteomics[100]. A recent study[101] leveraged machine learning to integrate immunofluorescence images from the Human Protein Atlas[102] with affinity purification data from BioPlex[103] to create a unified hierarchical map of human cell architecture, known as the multi-scale integrated cell (MuSIC 1.0). This map resolves 69 subcellular systems, paving the way for incorporating diverse types of data into proteome-wide cell maps. Furthermore, combining spatial proteomics-based tissue expansion with super-resolution and electron microscopy, this approach utilizes fluorescent labeling to achieve nanometer-scale resolution, enabling precise analysis of protein localization and interactions, while providing comprehensive subcellular-level insights into the spatial organization and functional networks of proteins.

However, integrating proteomics data with imaging remains a challenge.

i) Incompatibility between MS and antibody-based imaging. MS-proteomics is not directly compatible with antibody-based imaging methods like CODEX[104], multiplexed ion beam imaging[105], or 3D imaging mass cytometry[106], making integration difficult.

ii) Mismatch in data scale and coverage. Unlike spatial transcriptomics[107], which maps entire slides – albeit at great costs, MS-based methods by their nature focus on specific regions or small cell subsets guided via imaging[108], leading to challenges in data integration.

iii) Throughput limitations. MS-based spatial proteomics has lower throughput compared to transcriptomics, limiting large-scale and unbiased tissue mapping. However, by focusing on specific cells of interest in each slide, large cohorts can still be analyzed.

iv) Spatial proteomics dynamics. Spatial proteomics changes dynamically, requiring AI to identify patterns and predict trajectories that static methods cannot capture.

v) Limited spatial resolution. Some methods like microscaffold-assisted spatial proteomics have relatively low spatial resolution, which can be improved by AI. For example, a recent study leveraging deep learning to achieve high-resolution mapping of thousands of proteins in whole tissue sections [109].

Overcoming these obstacles will unlock the full potential of AI-driven spatial proteomics.

**Key area 5: perturbation proteomics**

Proteomics stands apart in its ability to capture dynamic cellular responses, offering significant advantages over static genomics and partially dynamic transcriptomics. Perturbation proteomics systematically investigates protein-level changes in response to various stimuli, providing critical insights into cellular systems and drug responses. This approach encompasses the comprehensive analysis of protein targets, interactions, PTMs, turnover, localization, and enzyme activities, alongside proteome quantification before and after perturbation[110].

Budding yeast is the first species where proteomes[111], transcriptomes[112], metabolomes[113], ionomes[114], as well as genetic interactions[115] became available for an extensively phenotypic library of all non-essential gene-knockouts[116]. Perturbation 'omics' has closed gaps in gene functional annotation, revealed general principles of the network responses, for instance in protein dynamics, and enabled the reconstruction of biological networks of unprecedented depth. The current frontier lies in leveraging this data through AI approaches to predict phenotypic responses, in the conditions in which they have been recorded, and across conditions and in environments not tested a *priori*.

Recent studies highlight an equally high potential for perturbation proteomics in human disease biology and drug discovery. Large-scale initiatives have profiled perturbation proteomes of numerous compounds, covering thousands of proteins and PTMs to elucidate for mechanisms of action and enable compound repurposing[117-120]. These studies have incorporated time and dose-resolved proteomics for a more nuanced, functional context to the observed changes[117-121].

Despite these advances, several challenges remain:

i) Limited data availability. Robust AI model training requires extensive perturbation data. Novel approaches, such as the TransPro model[122], generates synthetic perturbed proteome data from baseline transcriptome data, potentially bridging this gap.

ii) Strategic experimental design. Given the still high cost of proteomic experiments, developing strategies to generate limited yet highly informative perturbation data is crucial. While traditional experimental design relies on prior biological knowledge, AI algorithms could iteratively guide the selection of the most informative perturbations and biological models, creating a synergistic loop between experimentation and computation.

iii) Specialized AI algorithm development. To better understand proteomic dynamics, tailored AI algorithms are needed. These should incorporate temporal machine-learning techniques for comprehensive protein network analysis and multi-task predictions, accurately modeling the complex, time-dependent nature of cellular responses.

iv) Multi-omics integration. Leveraging transfer learning approaches and integrating data from various omics platforms can provide a more holistic view of cellular responses, potentially revealing emergent properties not visible in single-omics analyses.

**Key area 6: multi-omics data integration**

Advancements in high-throughput sequencing have catalyzed multi-omics (*e.g.*, genomics, transcriptomics, proteomics, and metabolomics) integration within the central dogma, enhancing disease research and precision medicine. Over the past two decades, databases like CPTAC, the UK Biobank, TCGA, ICGC, GEO, and LinkedOmics have accumulated multi-omics, phenotypic, and imaging data. These expanding population-scale biomedical datasets present unprecedented opportunities for AI to uncover complex biological links to health and disease.

Landmark studies demonstrate the power of this approach to uncover diagnostic biomarkers, potential drug targets, and risk stratification for various diseases. The MILTON machine learning model harnesses UK Biobank multi-omics and phenotype data to predict over 3,200 diseases[123]. Similarly, medical AI models trained on multimodal datasets successfully interpret diverse data types, including imaging, electronic health records, and genomics[124]. Recently, the Prov-GigaPath foundation model, trained on 1.3 billion pathology image tiles from a US health network spanning 28 cancer centers, improved cancer subtyping and mutation prediction, demonstrating AI's potential in clinical decision-making[125, 126].

However, significant challenges persist in integrating multi-omics data.

i) Data harmonization. Multi-omics integration is hindered by unpaired and/or discordant datasets, with varying sensitivities, specificities, coverage, and technical biases across platforms. Furthermore, different quantification methods, data scales, and preprocessing protocols further complicate comparisons between genes, transcripts, and proteins.

ii) Data format heterogeneity and limited accessibility. Heterogeneous storage formats, institutional data isolation, and data sparsity create barriers to integration. Additionally, multi-omics datasets suffer from noise and lack comprehensive databases, constraining AI model development.

iii) Current algorithmic constraints. While AI models like MIMaL and MOGONET have shown promising performance., approaches such as TMO-Net, MultiVI, and MOSA continue to struggle with in predicting missing omics data and integrating incomplete samples at scale.

iv) Challenges in single-cell multi-omics analysis. Single-cell multi-omics integration necessitates algorithms for vertical (cross-omics), horizontal (batch correction), and mosaic (datasets sharing at least one omics type) integration[127]. Recent models like scPROTEIN[128] tackle peptide quantification uncertainty, denoise protein data, and mitigate batch effects, but significant hurdles remain in linking SCP to functional outcomes.

v) Post-protein QC and functional interpretation. Beyond data integration, post-protein QC, accurate quantification, and establishing robust proteins disease associations remain critical barriers [129]. Looking ahead, future tools may process raw data end-to-end, enabling direct prediction of defined endpoints.

To advance the field, several key priorities emerge: generating paired multi-modal data, with particular emphasis on integrating more proteomics data, and developing models that bridge multi-omics with text or images. The integration of proteogenomic data promises to link genomic information with proteomic evidence, deepening our understanding of protein-coding regions and their functions. Moreover, the synthesis of spatial proteomics, SCP, PPIs, and imaging data within AI models will yield unprecedented insights into protein localization, cell-specific functions, and spatial interactions, advancing toward a more comprehensive view of proteomic landscapes.

**Key area 7: proteomics support to construct the AI virtual cell (AIVC)**

The concept of an AI virtual cell (AIVC) represents a computational model that simulates molecular, cellular, and tissue behavior, promising to accelerate discoveries and guide experiments[130]. Realizing this vision requires the seamless integration of multi-omics data (genome, epigenome, transcriptome, proteome, metabolome) alongside spatial, temporal, and imaging data[131].

To build AIVC, in addition to integrating the aforementioned proteomics and multi-omics data, we also need to annotate and predict protein functions. Despite numerous AI-based protein function annotation tools[132], predicting protein function remains challenging due to limited data and complex functional interactions[133]. A greater challenge is extracting functionally relevant insights from large datasets of proteins, PTMs, and their abundance changes across conditions. For instance, studies on the functional relevance of PTMs are essential. An ML model has been developed to predict functional scores for human phosphosites, using 59 features from large-scale data and databases like UniProt, helping prioritize sites for further study[134]. Similar approaches could be applied to prioritize other PTMs like acetylation, methylation, SUMOylation, and ubiquitination, although the lack of functional annotations for these PTMs limits model training. One solution is to leverage large language models to extract functional annotations for PTM sites from the literature[135]. Additionally, proteome-wide predicted protein structures offer new opportunities for genome-wide function prediction and have been successfully used to advance protein function understanding[136,137]. Interrogating PTMs in their structural context could further enhance our understanding of their functional relevance[138].

Within this framework, proteomics contributes multiple essential dimensions to our understanding of cellular biology: protein function prediction clarifies molecular roles, interaction data defines cellular networks, spatial proteomics ensures precise subcellular localization, and perturbation proteomics reveals dynamic responses. The challenges in each proteomics aspect and how AI can address them have been discussed in the previous sections. Successfully addressing these challenges will catalyze the development of AIVC.

The realization of a comprehensive digital cell requires large-scale proteomics and multi-modal data to capture the dynamics of molecular machinery and their networks, heralding a revolution in biological insights[131]. Progress in this field has followed a stepwise approach, beginning with whole-cell models for simple lower model organisms, such as the pathogen *Mycobacterium genitalium*[139] and the bacterium *Escherichia coli*[140]. For these simpler organisms present ideal initial targets: their fewer genes enable comprehensive analysis through existing sequencing technologies while traditional statistical algorithms may be sufficient to establish relatively reliable models. Current computational approaches span multiple frameworks, including ordinary differential equations (ODE)[141], Boolean network modeling[142], stochastic simulations[143], agent-based modeling[144], and constraint-based modeling[145].

The complexities of human cells, however, demand more sophisticated solutions. State of the art sequencing technologies and AI algorithms are essential for accurate modeling the behaviors of human cells. Addressing this challenge, the Human Cell Atlas (HCA) consortium was founded in 2016 to characterize all cell types in the human body[146]. This initiative has already collected single cell multi-omics data from 62 million cells, organized within 18 biological networks.

While most of the data derives from single cell RNA sequencing, its analysis leverages sophisticated neural network architectures, including transformer, convolutional neural network (CNN), Graph neural network (GNN), and diffusion model. This data can be used for building foundation models to extrapolate cell types and interaction networks[147], as exemplified by GeneFormer[148], and scFoundation[149].

Biological systems exist and function as communicating networks of as cells in dynamic environments, shaped by inter- and intra-cellular communication. Consequently, employing diverse perturbations to mimic various environmental effects on cells can enable deeper understanding of these intricate systems. Critical to this endeavor is the fusion of super-resolution imaging data with perturbation omics data which will enable a holistic view of cell dynamics for AI modeling. Tackling these challenges through interdisciplinary collaborations and advanced techniques will pave the way for future breakthroughs in predicting cellular behaviors, elucidating biological mechanisms, enabling personalized medicine, accelerating drug discovery, and pioneering new approaches in cellular engineering.

**Conclusion**

We identify seven key areas in AI-driven proteomics poised for breakthroughs within 3–5 years. At the foundation lies the development of an AI-friendly ecosystem for proteomics data, supporting six critical application areas: proteome identification and quantification (*e.g.*, canonical proteins, proteoforms, understudied proteins), protein complexes, PPIs, spatial proteomics, perturbation proteomics, and multi-omics integration.

Progress in these domains will collectively drive the AI virtual cell and achievement that demands global collaboration between AI and proteomics researchers, Initiatives like the π-Hub project [13] will serve as valuable catalyst in this scientific revolution.

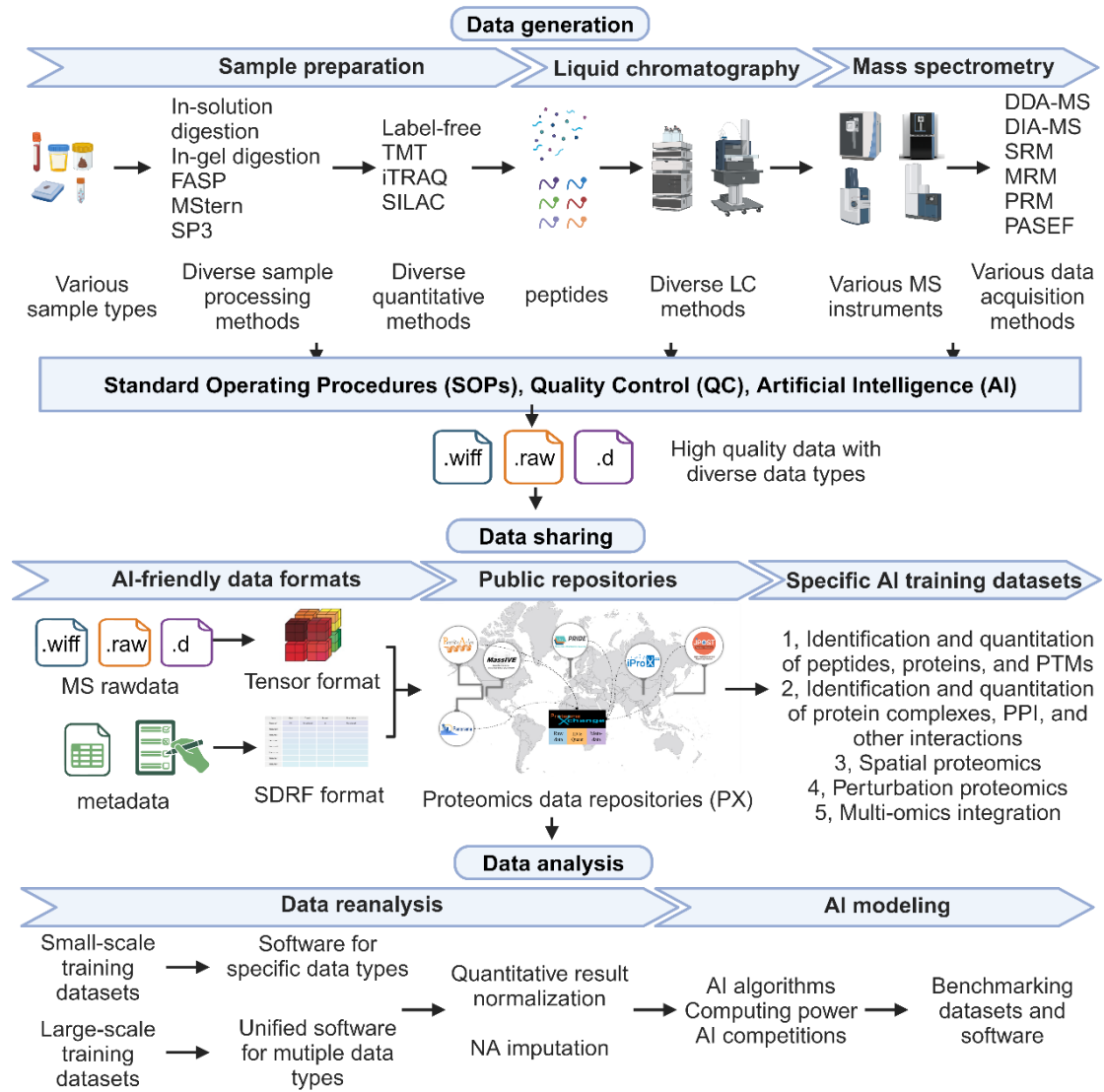

**Figure 1. Construction of AI-friendly ecosystem for proteomics data generation, sharing and analysis.**

The top, middle, and bottom sections illustrate how to establish an AI-friendly ecosystem for proteomics data across three stages: data generation, data sharing, and data analysis. For data generation, the figure outlines various sample types and proteomics methods used in sample preparation, liquid chromatography, and mass spectrometry. Each step requires standardized SOPs and QCs, along with AI integration to generate high-quality AI training datasets. In the data sharing section, diverse MS raw data types and metadata should be converted into AI-friendly formats, such as tensor and SDRF formats. These curated datasets can then be shared via public repositories like PX. Additionally, task-specific AI training datasets should be established for targeted downstream applications. For data analysis, appropriate specific and unified software should be selected based on dataset scale, with special attention to the standardization of quantitative results and handling of missing values. Finally, AI competitions should be organized to foster collaboration between AI and proteomics experts while generating benchmarking datasets to evaluate AI model performance.

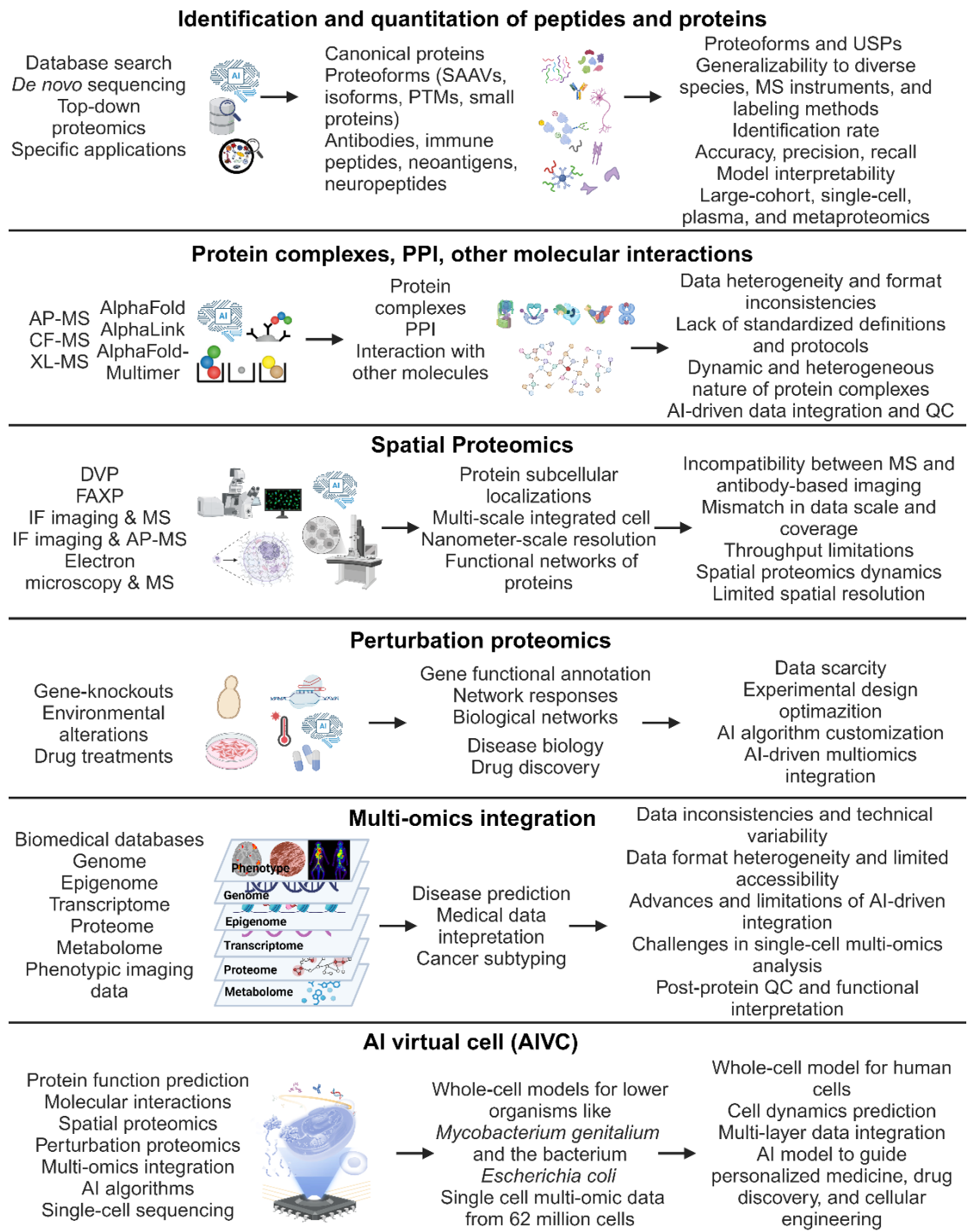

**Figure 2. Current advancements and challenges in each AI proteomics field.**

Six key areas currently being revolutionized by AI, including peptide and protein identification, protein complexes, PPI, molecular interactions, spatial proteomics, perturbation proteomics, multi-omics integration, and AI virtual cell. For each area, from left to right, the figure presents recent advancements, AI-driven achievements, and ongoing challenges.

**Box 1. Glossary**

| Full name | Abbreviation |
|---|---|
| Mass spectrometers | MS |
| Liquid chromatography | LC |
| Artificial intelligence | AI |
| Data-dependent acquisition | DDA |
| Data-independent acquisition | DIA |
| Tandem Mass Tag | TMT |
| Isobaric tags for relative and absolute quantitation | iTRAQ |
| Stable isotope labeling by amino acids in cell culture | SILAC |
| Top-down proteomics | TDP |
| Sequential window acquisition of all theoretical fragment ion mass spectra | SWATH |
| Parallel Accumulation–Serial Fragmentation | PASEF |
| Quality control | QC |
| Standard operating procedures | SOPs |
| Post-translation modifications | PTMs |
| Single amino acid variants | SAAVs |
| Protein-protein interactions | PPIs |
| Sample and data relationship format | SDRF |
| Peptide-to-spectrum match | PSM |
| AI virtual cell | AIVC |

**Box 2. Strategic Actions to Address Key Challenges in AI-Driven Proteomics**

**1, Development of an AI-friendly ecosystem for proteomics data**

Establish SOPs, QC methods, and AI algorithms to improve MS data quality.

Develop AI-friendly data formats and community pipelines for MS data reanalysis, normalization, and integration.

Generate synthetic data for benchmarking, algorithm evaluation, and experimental design optimization.

Create universal software and benchmarking datasets for large-scale AI training in proteomics.

Organize AI competitions to foster collaboration between AI and proteomics experts.

**2, Key area 2: Identification and quantitation of peptides and proteins**

Develop AI-driven methods to characterize proteoforms (e.g., SAAVs, isoforms, PTMs) and understudied proteins.

Enhance AI-based database search and de novo sequencing for diverse species, MS instruments, and labeling methods.

Improve the accuracy, identification rate, and interpretability of AI-driven proteomics tools.

Develop AI tools for TDP MS data interpretation and analysis.

Apply AI to advance large-cohort, single-cell, plasma, and metaproteomics research.

**3, Identification and quantification of protein complexes, protein-protein interactions, and other molecular interactions**

Develop AI-driven methods for data normalization, integration, and quality assessment of protein complex datasets.

Standardize formats, resolutions, and coverage to build AI-friendly, large-scale protein complex databases.

Use AI to model dynamic PPIs and protein complexes, capturing heterogeneity and mitigating noise from low-abundance complexes.

**4, Spatial proteomics**

Develop AI models to integrate MS-based proteomics with imaging data, addressing compatibility, scale, and throughput challenges.

Leverage AI for spatial proteomics to uncover dynamic patterns and predict molecular trajectories.

Enhance spatial resolution using deep learning for high-resolution protein mapping in tissues.

Optimize MS workflows and AI-driven data acquisition for higher efficiency and scalability.

**5, Perturbation proteomics**

Develop AI models for synthetic perturbation data generation to address data scarcity.

Optimize experimental design using AI, enabling the selection of the most informative perturbations and biological models.

Customize AI algorithms for proteomic dynamics, incorporating temporal machine-learning techniques for network analysis and multi-task predictions.

Leverage transfer learning for multi-omics integration, enhancing the holistic understanding of cellular responses.

**6, Multi-omics data integration**

Develop AI models to harmonize multi-omics datasets with varying sensitivities, specificities, and preprocessing methods.

Standardize storage formats and improve access to multi-omics datasets to enhance AI model development.

Optimize AI-driven models to predict missing omics data and integrate incomplete samples at scale.

Refine algorithms for vertical, horizontal, and mosaic integration to link single-cell proteomics with functional outcomes.

Integrate AI tools for post-protein quality control, accurate quantification, and linking proteins to diseases or phenotypes.

### 7, AI virtual cell (AIVC)

Use AI to integrate super-resolution imaging with perturbation omics for holistic cell dynamics modeling.

Develop AI models to predict cellular behavior and understand biological mechanisms under diverse perturbations.

Leverage AI to guide personalized medicine, drug discovery, and cellular engineering.

### Acknowledgements

T. G., Y. S., J. A., Z. L., R. S., L. Q., Y. C., Z. D., Y. D., and H. G. were supported by National Natural Science Foundation of China (Key Joint Research Program) (Grant No. U24A20476), National Natural Science Foundation of China (Major Research Plan) (Grant No. 92259201), National Natural Science Foundation of China (Young Scientist Fund) (Grant No. 32401239), Zhejiang Provincial Natural Science Foundation of China (LQ24C050002), and National Key R&D Program of China (Grant No. 2021YFA1301600). M. M. acknowledges Max Planck Society for the Advancement of Science. W. B. acknowledges the Research Foundation–Flanders (G087625N, G0AHY25N). C. L. was supported by an Australian Research Council (ARC) Future Fellowship (FT240100798) and a National Health and Medical Research Council of Australia (NHMRC) Ideas Grant (2024/GNT2037597). C. C. and F. H. were supported by National Key Research and Development Program of China (2024YFA1210400 and 2021YFA1301603), and the National Natural Science Foundation of China (32088101). J. A. V. and Y. P. R. acknowledge BBSRC grants BB/X001911/1, BB/V018779/1, APP9749, and Wellcome grant 223745/Z/21/Z. S. H. P. was supported by NIGMS/NIH award R01GM147653. H. H. acknowledges BBSRC grant BB/X002179/1. V.D. supported by BMBF grant 161L0221. M. L. acknowledges the National Key R&D Program of China (No. 2022YFA1304603) and Canadian NSERC Grant OGP0046506.

### Conflict of interest

T. G. is the founder of Westlake Omics Inc. M. M. is an indirect investor in Evosep. X. L. has a project contract with Bioinformatics Solutions Inc. J. R. K. is an employee of Bruker Ltd. Milton, Canada. Y.X. is an employee of Thermo Fisher Scientific. B.B.S. am currently a full-time employee of Bristol Myers Squibb. V. D. holds shares of Aptila Biotech. The remaining authors have not claimed conflict of interest related to this manuscript.